\documentclass[superscriptaddress,showpacs,preprintnumbers]{revtex4}
\usepackage{graphicx}
\usepackage{dcolumn}
\usepackage{bm}
\usepackage{natbib}

\input{tcilatex}

\begin{document}

\title{Role of interband scattering in neutron irradiated MgB$_2$ 
thin films by Scanning Tunneling Spectroscopy measurements}
\author{R. Di Capua}
\email{rdicapua@na.infn.it}
\affiliation{University of Napoli and CNR/INFM-Coherentia, Via Cinthia, I-80126 Napoli, Italy}
\author{H. U. Aebersold}
\affiliation{Paul Sherrer Institut, CH-5232 Villagen, Switzerland}
\author{C. Ferdeghini} 
\affiliation{CNR/INFM-LAMIA, Via Dodecaneso 33, I-16146 Genova, Italy}
\author{V. Ferrando}
\affiliation{CNR/INFM-LAMIA, Via Dodecaneso 33, I-16146 Genova, Italy}
\affiliation{The Pennsylvania State University, University Park, PA-16802, USA}
\author{P. Orgiani}
\affiliation{The Pennsylvania State University, University Park, PA-16802, USA}
\affiliation{CNR/INFM-SUPERMAT, I-84081 Baronissi (SA), Italy}
\author{M. Putti}
\affiliation{CNR/INFM-LAMIA, Via Dodecaneso 33, I-16146 Genova, Italy}
\author{M. Salluzzo}
\affiliation{University of Napoli and CNR/INFM-Coherentia, Via Cinthia, I-80126 Napoli, Italy}
\author{R. Vaglio}
\affiliation{University of Napoli and CNR/INFM-Coherentia, Via Cinthia, I-80126 Napoli, Italy}
\author{X. X. Xi}
\affiliation{The Pennsylvania State University, University Park, PA-16802, USA}


\begin{abstract}
A series of MgB$_2$ thin films systematically disordered by neutron
irradiation have been studied by Scanning Tunneling Spectroscopy. The c-axis
orientation of the films allowed a reliable determination of local density
of state of the $\pi$ band. With increasing disorder, the conductance peak 
moves towards higher voltages and becomes lower and broader, indicating a
monotonic increase of the $\pi$ gap and of the broadening parameter. These
results are discussed in the frame of two-band superconductivity.
\end{abstract}

\maketitle


The discovery of two gap superconductivity in MgB$_{2}$ \cite{Giu01,Iav02}
has renewed the interest on this mechanism that was theoretically predicted
since the fifties. This peculiar feature occurs in MgB$_{2}$ because of the
presence of two sets of bands which cross the Fermi level. The larger gap 
($\Delta $$_{\sigma }\sim$ 7 meV) is related to the $\sigma $ bands, originated by 
p$_{xy}$ B orbitals, strongly interacting with phonons, and characterized by a 
bidimensional Fermi surface; the smaller gap ($\Delta $$_{\pi }\sim$ 2 meV) is 
related to the $\pi $ bands originated by p$_{z}$ B orbitals and having an 
isotropic Fermi surface. 
Two band models predict that non-magnetic scattering causes pair breaking as 
magnetic one does in a one band superconductor;\cite{Gol97} this is a unique 
effect of interband scattering, because mixing $\sigma$ and $\pi$ Cooper's pairs 
causes a complete isotropization of the entire Fermi surface. In the strong 
interband scattering limit, the two gaps should merge to one when the 
critical temperature (T$_c$) drops to the isotropic value of about 
20-25 K.\cite{Liu01,Cho02} To verify these predictions, 
several efforts have been made to evaluate the energy gaps in samples where 
defects were introduced by different techniques such as substitutions (Al 
\cite{Put03,Put05,Kar05} or C \cite{Sam03,Sch03,Hol04,Gon05}), irradiation 
\cite{Wan03,Put06} or in films grown naturally disordered.\cite{Iav05} Due to
the possible simultaneous occurrence of several sources of disorder, a
quantitative evidence of the role of the interband scattering has not yet
been given. In particular, substitution may induce extrinsic effects related
to charge doping, structural instability and inhomogeneous distribution of
impurities. However, it is widely accepted that the $\Delta $$_{\pi }$ value
is weakly affected by disorder while $\Delta $$_{\sigma }$ decreases
linearly with T$_{c}$, and recently the merging of the gap has been observed
in irradiated samples at a critical temperature (11 K) lower
than the predicted one.\cite{Put06}

In this paper, we report on Scanning Tunneling Spectroscopy (STS)
measurements on c-oriented thin films where the defects were systematically
introduced by neutron irradiation. Neutron irradiation introduces
neither charge doping nor electronic or structural instabilities;
furthermore, in thin films it guarantees a uniform distribution of defects 
in the entire sample.\cite{Fer06} STS directly probes the Local Density of States
(LDOS), whose broadening can be analyzed in connection with scattering
mechanisms induced by irradiation.

MgB$_{2}$ thin films (2000 $\AA$ thick) were grown at The Pennsylvania State University 
by Hybrid Physical Chemical Vapor Deposition (HPCVD) on 5x5 mm$^{2}$ SiC substrates. Details 
about the samples fabrication and their basic characterization can be found 
in Ref. \onlinecite{Zen02}. Neutron irradiation was carried out at the spallation neutron source 
SINQ of Paul Sherrer Institut in Villigen (Zurich). Each film was sealed under vacuum 
in a small quartz ampoule just after deposition; this setup allowed to perform
irradiation also for long times without compromising samples quality. 
Several samples (Samples IRR10, IRR30, IRR35, and IRR40 in the following) 
have been irradiated at thermal neutron fluences ranging from 6.4$\cdot$10$^{15}$ 
to 9.5$\cdot$10$^{18}$ cm$^{-2}$. A description of the damage mechanism, as well as 
detailed structural, transport and magnetic characterization were reported elsewhere.\cite{Fer06} 
Low temperature STS experiments were performed in a cryogenic
system able to operate at variable temperature and in magnetic field, using
a PtIr home-made tip. The films were mounted on a Scanning Tunneling
Microscope scanning head in inert helium atmosphere to preserve the surface
quality.

Table~\ref{tab:properties} reports the main features of the measured samples: 
as the fluence increases, T$_c$ value changes from 41 K to 17 K, and the residual 
resistivity $\rho$$_0$ varies by two orders of magnitude. 

\begin{table}[tbp]
\caption{Main properties of the measured irradiated films.}
\label{tab:properties}%
\begin{ruledtabular}
\begin{tabular}{cccc}
Sample&Neutron Fluence (cm$^{-2}$)&T$_c$(K)&$\rho$$_0$ ($\mu$$\Omega$$\cdot$cm)\\
\hline
IRR10 & 6.4$\cdot$10$^{15}$ & 41.05 & 1.2\\
IRR30 & 7.7$\cdot$10$^{17}$ & 36.1 & 18\\
IRR35 & 3.0$\cdot$10$^{18}$ & 22.2 & 52\\
IRR40 & 9.5$\cdot$10$^{18}$ & 17.0 & 87\\
\end{tabular}
\end{ruledtabular}
\end{table}

\begin{figure}[tbp]
\includegraphics[width=12cm,height=8.62cm, bb = 0 0 826 594]{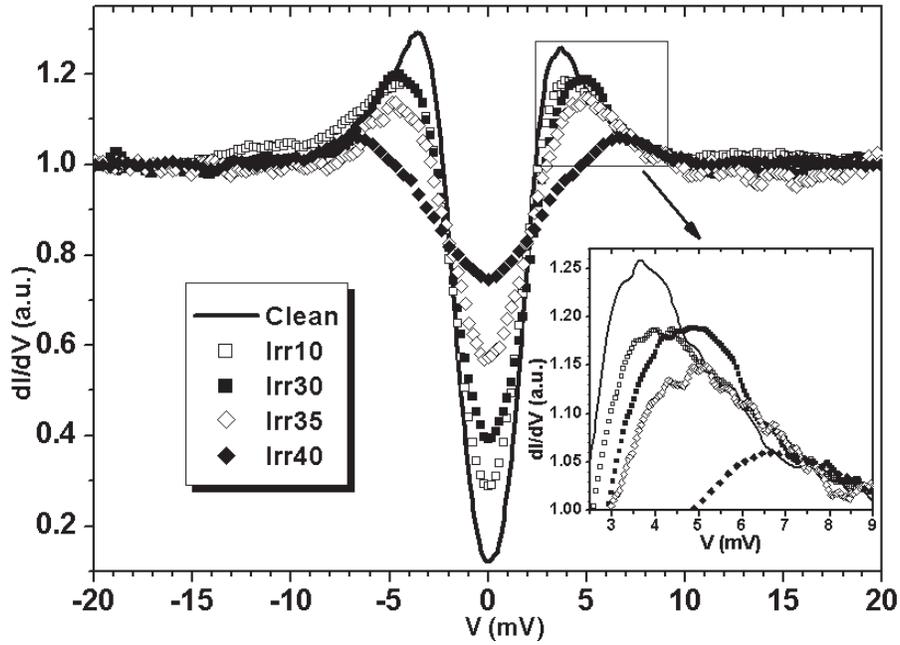}
\caption{STS spectra, at T = 4.2 K, on the irradiated measured MgB$_2$ films. 
Measurements parameters: tunnel current: 100 pA; bias voltage: 20 mV. 
A spectrum on a clean sample, from Ref. \onlinecite{Iav05}, is also reported as reference.}
\label{fig:SpectraComparison}
\end{figure}

Figure~\ref{fig:SpectraComparison} shows the typical normalized tunneling conductance 
STS spectra for neutron irradiated films of different fluences. 
All the spectra were collected through a standard lock-in technique and 
were acquired (at T=4.2 K) by stabilizing the feedback loop with a tunnel 
current of 100 pA and a bias voltage of 20 mV. 
The STS spectra on each sample were reproducible from 
point to point on the sample surface, and they overlapped when normalized by
changing the tunnel resistance. This assures that we were measuring in pure tunneling 
regime and guarantees a reduced role of possible contaminated surface layers. 
STS measurements were performed only when such conditions were achieved. A 
spectrum on a non irradiated film (T$_c$ = 41 K, from Ref. \onlinecite{Iav05}), measured 
in the same conditions, is also reported as reference.

The plot reveals a clear monotone trend of the spectra. As T$_c$ decreases, 
with increasing neutron fluences, the zero-bias conductance (ZBC) increases 
and the coherence superconductivity peaks shift to higher voltages and appear 
less pronounced. This behavior is more clearly shown in the inset to Fig.~\ref{fig:SpectraComparison}. 
Only one-peak spectra were found on all the samples. Because in c-axis oriented 
films the current is injected parallel to the c-axis, the $\pi$ band contribution 
is dominant in the tunneling spectra as compared to the $\sigma$ band 
contribution.\cite{Bri02} Therefore, we identify the observed peaks as corresponding 
to the $\pi$ gap, $\Delta$$_{\pi}$.

The increase of the ZBC is likely due to a broadening of the superconducting LDOS, 
as a consequence of the increasing disorder, which introduces subgap states 
and smears the LDOS divergence. The displacement of the peaks towards higher 
energies could be ascribed mainly to two effects: the LDOS broadening, supported 
by the simultaneous decrease in the peak height, and an intrinsic change of 
the $\Delta$$_{\pi}$ value as disorder increases and T$_c$ decreases.

\begin{figure}[tbp]
\includegraphics[width=12cm,height=6.7cm, bb = 0 0 750 425]{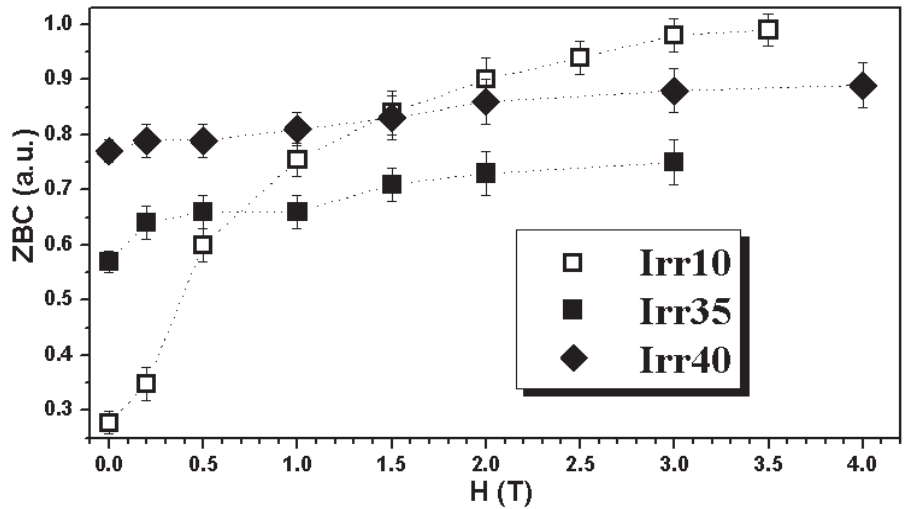}
\caption{Zero Bias Conductance as a function of the applied magnetic field (perpendicular 
to the surface) for samples Irr10, Irr35, Irr40.}
\label{fig:ZBCvsH}
\end{figure}

Measurements in magnetic field perpendicular to the films surface were also performed. 
Figure~\ref{fig:ZBCvsH} reports the ZBC as a function of the applied field for 
some samples. The plot shows that the most irradiated samples are much less sensitive 
to the magnetic field, as already observed in differently disordered samples.\cite{Iav05} 
This result will be discussed in detail elsewhere.

\begin{figure}[tbp]
\includegraphics[width=12cm,height=8.61cm, bb = 0 0 804 577]{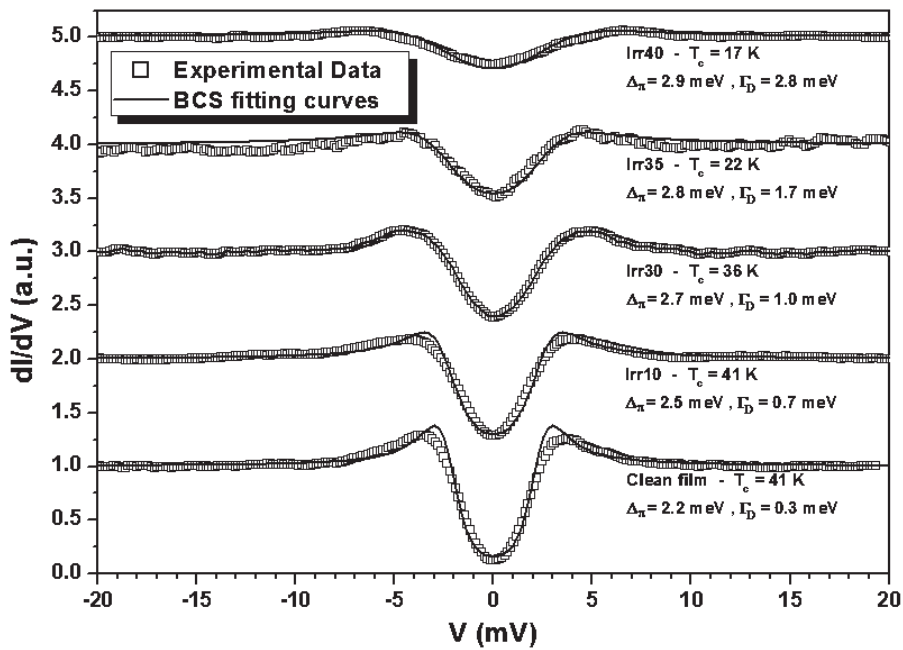}
\caption{STS spectra, at T = 4.2 K, on the irradiated measured MgB$_2$ films; 
fitting curves, evaluated by a BCS one-gap model, are superimposed to the experimental 
data. Plots are vertically shifted for clarity.}
\label{fig:DataAndFits}
\end{figure}

To get further information on the evolution of LDOS as a function of 
irradiation induced disorder, a quantitative analysis can be developed. As a 
first approximation, we fitted the measured dI/dV spectra through a simple 
(single gap) BCS model with a phenomenological Dynes parameter $\Gamma$$_D$.\cite{Dyn78} 
In fact, when only the $\pi$ band gives a significant 
contribution to the tunnel current, the two-band calculation for MgB$_2$ 
approaches a simple one-band BCS model,\cite{Bri02} with $\Delta$$_{\pi}$ and 
$\Gamma$$_D$ as fitting parameters. The $\Gamma$$_D$ parameter includes all 
the measured broadening on the superconducting LDOS, due to intrinsic, 
pair-breaking, effects, but also to extrinsic effects such as thermal noise. 
The results of the best fit curves and fitting parameters are shown 
in Fig.~\ref{fig:DataAndFits}.

The good agreement between the fits and data in Fig.~\ref{fig:DataAndFits} confirms the 
adequacy of the one-band model. Although the high $\Gamma$$_D$ value estimated for the 
lowest T$_c$ sample seriously affects the quantitative reliability of the simple model, 
it furnishes anyway an indication of the film behavior. The use of a two band model, with 
the estimated $\sigma$ gap from the specific heat data on bulk samples 
irradiated through the same technique,\cite{Put06} does not change the 
extracted $\pi$ gap value substantially, but introduces more free parameters, 
making the overall analysis less straightforward and clear.

\begin{figure}[tbp]
\includegraphics[width=12.6cm,height=8.35cm, bb = 0 0 802 532]{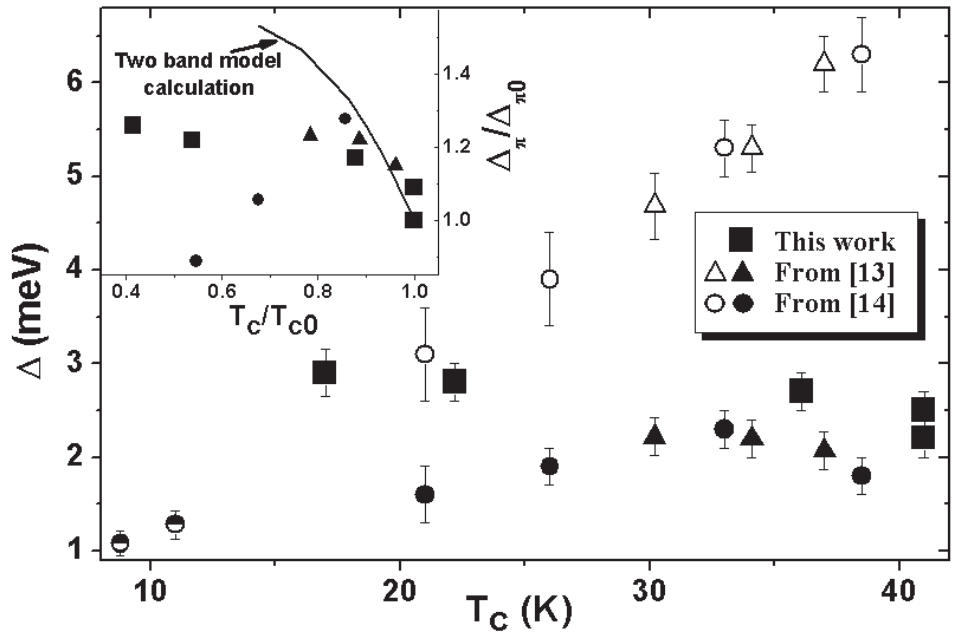}
\caption{$\Delta$$_{\pi}$ values evaluated by STS spectra as a function of T$_c$ (squares). 
For comparison, $\Delta$$_{\pi}$ and $\Delta$$_{\sigma}$ evaluated by specific heat measurements 
on neutron irradiated polycrystalline samples (triangles, from Ref. \onlinecite{Wan03}, and circles, from 
Ref. \onlinecite{Put06}) are also reported. In the inset, $\Delta$$_{\pi}$/$\Delta$$_{\pi}$$_0$ vs. T$_c$/T$_{c0}$ 
is plotted together with a theoretical calculation.}
\label{fig:GapsVsTc}
\end{figure}

Figure~\ref{fig:GapsVsTc} shows $\Delta$$_{\pi}$ values as a function of T$_c$ measured from 
our STS spectra. $\Delta$$_{\sigma}$ and $\Delta$$_{\pi}$ evaluated by specific heat 
measurements on neutron irradiated polycrystalline samples \cite{Wan03,Put06} 
are also plotted for comparison. The $\Delta$$_{\pi}$ values estimated by STS are usually 
reported to be larger than those estimated by specific heat. This accounts for the slightly 
different $\Delta$$_{\pi}$ values obtained by the different measurements for the lightly 
irradiated samples. The $\Delta$$_{\pi}$ value estimated by STS, even considering the 
error bars, clearly increases as T$_c$ decreases. In heavily irradiated thin films 
(T$_c$ = 22 and 17 K), the extracted $\Delta$$_{\pi}$ value from the STS spectra 
are close to $\Delta$$_{\sigma}$ estimated by specific heat.

The increase of $\Delta$$_{\pi}$ is predicted by the two-band model for increasing interband scattering.\cite{Gol97,Dol05} 
Up to now, an experimental proof of this effect is lacking due 
to the difficulty to introduce disorder in a controlled way without simultaneous doping effects. 
With the reliable gap data from the neutron irradiated samples a comparison with the theory 
is possible. In the inset to Fig.~\ref{fig:GapsVsTc}, $\Delta$$_{\pi}$/$\Delta$$_{\pi}$$_0$ data as 
a function of T$_c$/T$_{c0}$ from this work and from Refs. \onlinecite{Wan03} and \onlinecite{Put06} are shown, where 
$\Delta$$_{\pi}$$_0$ and T$_{c0}$ are the $\pi$ gap and the T$_c$ values of the 
unirradiated samples, respectively. The solid line represents the theoretical prediction 
of the two band model for increasing interband scattering.\cite{Dol05} 
For T$_c$/T$_{c0}$ ranging from 1 to 0.85 ($\Delta$T$_c$ = T$_{c0}$–T$_c$ $\sim$ 6 K), 
all the three data series show a similar increase of $\Delta$$_{\pi}$/$\Delta$$_{\pi}$$_0$ in agreement 
with the theoretical curve. This suggests that for low levels of disorder the main mechanism 
of T$_c$ reduction is the pair breaking due to interband scattering. In the regime of small interband 
scattering, the scattering rate, $\Gamma$$_{inter}$, can be calculated  as \cite{Maz03} 
$\Delta$T$_c$/T$_{c0}$ = 0.2$\Gamma$$_{inter}$/k$_B$T$_{c0}$; with $\Delta$T$_c$ = 6 K we estimate 
$\Gamma$$_{inter}$ $\sim$ 2.5 meV.

For T$_c$/T$_{c0}$ < 0.85 the experimental data do not agree; the 
$\Delta$$_{\pi}$/$\Delta$$_{\pi}$$_0$ values of this work are nearly constant and those from the 
specific heat on bulk samples decrease with decreasing T$_c$. In both cases, the data remain below the theoretical 
curve that continues to increase. Thus for these levels of disorder other mechanisms cause the decreasing 
of $\Delta$$_{\pi}$, and consequently the suppression of the T$_c$. Also for strong interband 
scattering $\sigma$ and $\pi$ Cooper pairs are expected to mix, but the single-gap model used 
to extract $\Delta$$_{\pi}$ considers only the $\pi$ contribution. 
Not surprisingly, for the sample with T$_c$=17 K we found that 2$\Delta$$_{\pi}$/k$_B$T$_c$ = 3.9, 
which is too high for the $\pi$ band but reasonable for the $\sigma$ band.

\begin{figure}[tbp]
\includegraphics[width=12cm,height=8.55cm, bb = 0 0 660 470]{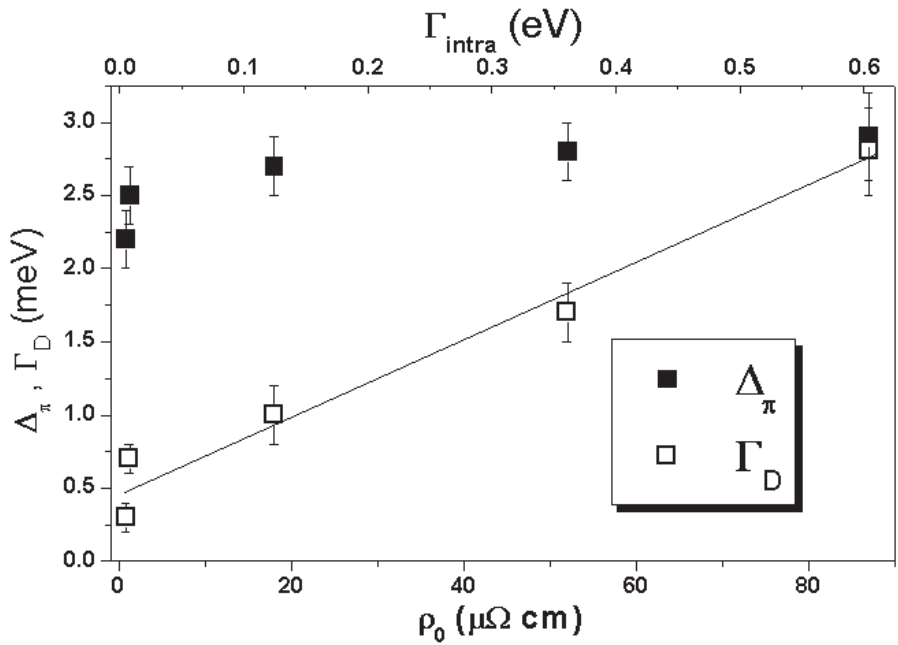}
\caption{$\Delta$$_{\pi}$ and $\Gamma$$_D$ values estimated by tunnel spectra as a function or residual 
resistivity $\rho$$_0$ (lower scale). On the upper x-scale, the intraband scattering $\Gamma$$_{intra}$ 
has been plotted, estimated as in Ref. \onlinecite{Pal05} in the hypothesis of equal relaxation rate 
in $\sigma$ and $\pi$ bands. The continuous line is only a guide for eyes.}
\label{fig:GapAndGamVsRho}
\end{figure}

In Fig.~\ref{fig:GapAndGamVsRho} the estimated Dynes broadening parameter $\Gamma$$_D$ is reported as a function 
of $\rho$$_0$. It increases linearly with $\rho$$_0$ from 0.3 to 2.8 meV. This linear correlation seems to suggest 
that quasiparticle relaxation processes, which affect $\rho$$_0$, could enhance pair breaking mechanisms which 
influence $\Gamma$$_D$. A linear increase of $\Gamma$$_D$ with resistivity was first observed and discussed in 
Ref. \onlinecite{Dyn84}, which suggested that finite lifetime effects due to inelastic electron-electron scattering 
vary linearly with resistivity, being enhanced by disorder. In a two gap superconductor, in principle both 
intraband and interband events can contribute to the observed scattering processes, which 
cannot be sorted out from the present data. However, the intraband relaxation rate, $\Gamma$$_{intra}$, can be 
roughly estimated by $\rho$$_0$ assuming equal intraband relaxation rate in $\sigma$ and $\pi$ bands.\cite{Pal05} 
It ranges from 0.1 to 0.6 eV (see upper scale in Fig.~\ref{fig:GapAndGamVsRho}), much higher than $\Gamma$$_D$. 
Interestingly $\Gamma$$_D$ values are close to $\Gamma$$_{inter}$. For example, the sample with T$_c$ = 36K has 
$\Gamma$$_{inter}$ $\sim$ 1.6meV and $\Gamma$$_D$ = 1meV. We believe that this 
agreement is not fortuitous. Indeed, in a two gap superconductor the interband scattering is expected to cause 
pair breaking as scattering with magnetic impurities does in single gap superconductor; 
therefore, it is reasonable that the interband scattering can strongly influence the broadening of the 
tunnel LDOS of the $\pi$ band. Furthermore, this framework implies a linear dependence between interband and 
intraband scattering rates (suggested by the linear dependence of $\Gamma$$_D$ on $\rho$$_0$, as discussed above) 
as both mechanisms scale with the defect density, which increases with 
the neutron fluence. Since the extrapolated low resistivity $\Gamma$$_D$ value is different 
from zero, part of the LDOS broadening should be ascribed to extrinsic factors such as finite 
noise temperature or presence of a contaminated surface layer.

In conclusion, we measured, by STS, neutron irradiated 
MgB$_2$ thin films grown by HPCVD. From tunnel spectra 
analysis we estimated the behavior of $\Delta$$_{\pi}$ 
and $\Gamma$$_D$as a function of the introduced disorder. 
At low level of disorder (T$_c$ lowered from 40 to 36 K) 
$\Delta$$_{\pi}$ increases slightly as an effect of interband 
scattering, in quantitative agreement with the two band theory. 
Furthermore, by comparing the behavior of the $\Gamma$$_D$ 
parameter with resistivity we were also able to discuss the 
scattering mechanisms and reasonably 
infer the role of the interband scattering in 
determining the broadening of the measured density of states.

This work is partially supported by Ministry of Italian Research 
by PRIN2004022024 project. The work at Penn State is supported in 
part by NSF under Grant No. DMR-0306746 and by ONR under Grant 
No. N00014-00-1-0294.

\end{document}